\documentstyle[seceq]{ptptex}

\title{
Algebraic Properties of the Real Quintic Equation 
for a Binary Gravitational Lens
}

\author{
Hideki {\sc Asada},$^{1,2,}$\footnote{E-mail: 
asada@phys.hirosaki-u.ac.jp} 
Taketoshi {\sc Kasai}$^{1}$ 
and 
Masumi {\sc Kasai}$^{1,}$\footnote{E-mail: 
kasai@phys.hirosaki-u.ac.jp}
}

\inst{
$^1$ 
Faculty of Science and Technology, Hirosaki University, 
Hirosaki 036-8561, Japan\\
$^2$ 
Max-Planck-Institut f\"ur Astrophysik, 
Karl-Schwarzschild-Str. 1 \\
D-85741 Garching, Germany
}

\recdate{
}

\abst{
It has been recently shown that the lens equation for a binary 
gravitational lens, which is apparently a coupled system, 
can be reduced to a real fifth-order (quintic) algebraic equation. 
Some algebraic properties of the real quintic equation are revealed. 
We find that the number of images on each side of the separation axis 
is independent of the mass ratio and separation 
unless the source crosses the caustics. 
Furthermore, the discriminant of the quintic equation enables us 
to study changes in the number of solutions, namely 
in the number of images. 
It is shown that this discriminant can be factorized into two parts: 
One represents the condition that the lens equation can be reduced to 
a single quintic equation, while the other corresponds to the caustics. 
}

\begin{document}

\maketitle

\section{Introduction}
Gravitational lensing by binary objects has attracted  
a great deal of interest \cite{SEF} since the pioneering work of  
Schneider and Wei{\ss} (1986).\cite{SW} It offers an important 
tool in the search for extrasolar planets,\cite{MP,GL} which is 
independent of the Doppler method, which suffers from uncertainties in 
inclination angles. \cite{MB}  
However, it is not easy to study binary lensing in detail: 
The lens equation adopted until now consists of a set of 
two coupled real fifth-order (quintic) equations, or, equivalently, 
a complex quintic equation \cite{Witt90} that is expressed in 
a complex notation. \cite{BKN,BK} 
As a result, it seems quite difficult to solve these equations, 
since there are no well-established method for solving coupled 
nonlinear equations numerically with sufficient accuracy. \cite{NR} 

For a symmetric binary with two equal masses, the lens equation 
for a source on the symmetry axes of the binary becomes 
so simple that we can find analytic solutions. \cite{SW} 
For star-planet systems, furthermore, the mass ratio of the binary 
is so small that we can find approximate solutions in general.
\cite{Bozza,Asada02a} Such approximate solutions have been used to 
study the shift of the photocenter position due to astrometric 
microlensing. \cite{Asada02a}  

One of the present authors recently showed that the lens equation can 
be reduced to a single master equation that is fifth-order 
in a real variable with real coefficients.\cite{Asada02b} 
As a consequence, the new formalism provides a much simpler method  
to study binary lensing. 
In this paper, we reveal algebraic properties of the real quintic
equation that have not yet been studied. \cite{Asada02b} 
The discriminant of the quintic equation is shown to be factorizable 
into two parts: One represents the condition that the lens equation 
can be reduced to a single quintic equation, while the other 
corresponds to the caustics. 

The caustics are curves on which the Jacobian of the lens mapping 
vanishes on a source plane. 
The number of images has been classified in terms of caustics. 
Caustics for two-point masses have been investigated in detail 
on the basis of a set of two coupled real fifth-order equations 
as an application of catastrophe theory \cite{ES} 
and with a complex formalism.\cite{WP} 
In binary lensing, three images appear for a source 
outside the caustics, while five images appear for a source 
inside the caustics. 
Regarding the number of images, it should be straightforward to 
count the number of solutions for the lens equation. 
However, this has not been done for a binary lens system,  
because the lens equations used are coupled nonlinearly. 
One of the main purposes of the present paper is to count 
these solutions. 

This paper is organized as follows. 
First, the derivation of the real quintic lens equation is 
briefly summarized. 
Then, a condition for its reduction is presented. 
Next, the discriminant of the equation is computed, and in this way 
we clarify the relation between the discriminant and the caustics. 

\section{Discriminant of a real quintic lens equation} 
\subsection{A real quintic lens equation}
We consider a binary system of two point masses 
$M_1$ and $M_2$ and a separation vector $\mbox{\boldmath $L$}$ 
from object 1 to 2, which is located a distance 
$D_{\mbox{L}}$ from the observer. 
In units of the Einstein ring radius angle 
$\theta_{\mbox{E}}$, the lens equation reads 
\begin{equation}
\mbox{\boldmath $\beta$}=\mbox{\boldmath $\theta$}
-\Bigl( 
\nu_1 \frac{\mbox{\boldmath $\theta$}}{|\mbox{\boldmath $\theta$}|^2} 
+\nu_2 \frac{\mbox{\boldmath $\theta$}
-\mbox{\boldmath $\ell$}}{|\mbox{\boldmath $\theta$}
-\mbox{\boldmath $\ell$}|^2} 
\Bigr) , 
\label{lenseq}
\end{equation}
where $\mbox{\boldmath $\beta$}$ and $\mbox{\boldmath $\theta$}$ 
denote the vectors representing the positions of the source and image, 
respectively, and we defined the mass ratios $\nu_1$ and $\nu_2$ and 
the angular separation vector $\mbox{\boldmath $\ell$}$ as 
\begin{eqnarray}
\nu_1&=&\frac{M_1}{M_1+M_2} , \\
\nu_2&=&\frac{M_2}{M_1+M_2} , \\
\mbox{\boldmath $\ell$}&=&
\frac{\mbox{\boldmath $L$}}{D_{\mbox{L}}\theta_{\mbox{E}}} . 
\end{eqnarray} 
We have the identity $\nu_1+\nu_2=1$. 
For brevity, $\nu_2$ is denoted by $\nu$. 
Equation (\ref{lenseq}) is a set of two coupled real quintic
equations for $(\theta_x, \theta_y)$, equivalent to a single complex 
quintic equation for $\theta_x+i \theta_y$.\cite{Witt90,Witt93} 

Let us introduce polar coordinates whose origin is located at 
the position of object 1 and whose angle is measured from 
the separation axis of the binary. 
The coordinates for the source, image and separation vector are 
$(\beta_x, \beta_y)=(\rho\cos\varphi, \rho\sin\varphi)$, 
$(\theta_x, \theta_y)=(r\cos\phi, r\sin\phi)$ and 
$(\ell_x, \ell_y)=(\ell, 0)$, respectively, 
where $\rho$, $r$ and $\ell \geq 0$. 
For brevity, $\cos\varphi$ and $\sin\varphi$ are written as 
$C$ and $S$, respectively. The parallel and vertical parts of 
Eq. (\ref{lenseq}) with respect to the separation vector 
$\mbox{\boldmath $\ell$}$ are 
\begin{eqnarray}
&&r\cos\phi\Bigl(1-\frac{1-\nu}{r^2}-
\frac{\nu}{r^2-2\ell r\cos\phi+\ell^2}\Bigr) 
=\rho C-\frac{\nu\ell}{r^2-2\ell r\cos\phi+\ell^2} , 
\label{parallel}\\
&&r\sin\phi\Bigl(1-\frac{1-\nu}{r^2}-
\frac{\nu}{r^2-2\ell r\cos\phi+\ell^2}\Bigr)
=\rho S . 
\label{vertical} 
\end{eqnarray}

Two particular cases $\rho=0$ (for which the source is located 
behind the lens object $M_1$) and $S=0$ (for which the source is
located on the axis) have been studied,\cite{Asada02b} 
where analytic solutions for an arbitrary mass 
ratio were obtained explicitly. 
In the following, we concentrate on the case $\rho\neq 0$ and 
$S \neq 0$ (i.e. off-axis sources). 
In this case, it has been shown \cite{Asada02b} that 
Eq. ($\ref{lenseq}$) reduces to a fifth-order equation for $\cot\phi$, 
\begin{equation}
\sum_{i=0}^5a_i(\cot\phi)^{5-i}=0 , 
\label{theta2}
\end{equation}
where we have defined 
\begin{eqnarray}
a_0&=&\nu\ell\rho^3S^3 , 
\\
a_1&=&\rho^2S^2+2\ell\rho^3CS^2
-\ell^2(2\rho^2S^2-\rho^4S^2)
-2\ell^3\rho^3CS^2+\ell^4\rho^2S^2
\nonumber\\ 
&&
-\nu(5\ell\rho^3CS^2-4\ell^2\rho^2S^2) , 
\\
a_2&=&-2\rho^2CS-4\ell\rho^3(C^2S-S^3) 
+\ell^2(4\rho^2CS-2\rho^4CS)
+4\ell^3\rho^3C^2S-2\ell^4\rho^2CS 
\nonumber\\
&&+\nu\Bigl[\ell(2\rho S+8\rho^3C^2S-2\rho^3S^3) 
-10\ell^2\rho^2CS+2\ell^3\rho S\Bigr] , 
\\
a_3&=&\rho^2+\ell(2\rho^3C^3-10\rho^3CS^2)
+\ell^2(-2\rho^2C^2+2\rho^2S^2+\rho^4) 
-2\ell^3\rho^3C+\ell^4\rho^2 \nonumber\\
&&+\nu\Bigl[\ell(-2\rho C-4\rho^3C^3+6\rho^3CS^2) 
+6\ell^2\rho^2C^2-2\ell^3\rho C\Bigr]  
+\nu^2\ell^2 , 
\\
a_4&=&-2\rho^2CS+8\ell\rho^3C^2S-\ell^2(4\rho^2CS+2\rho^4CS)
+4\ell^3\rho^3C^2S-2\ell^4\rho^2CS 
\nonumber\\
&&+\nu\Bigl[\ell(2\rho S-4\rho^3C^2S+\rho^3S^3)
-2\ell^2\rho^2CS+2\ell^3\rho S\Bigr] , 
\\
a_5&=&\rho^2C^2-2\ell\rho^3C^3+\ell^2(2\rho^2C^2+\rho^4C^2)
-2\ell^3\rho^3C^3+\ell^4\rho^2C^2 
\nonumber\\
&&
+\nu\Bigl[-\ell(2\rho C+\rho^3CS^2) 
+2\ell^2\rho^2C^2-2\ell^3\rho C\Bigr]  
+\nu^2\ell^2 .  
\end{eqnarray}
Here, none of these coefficients $a_0, \cdots a_5$ are singular, 
because they are polynomials in $\ell$, $\rho$, 
$\nu$, $C$ and $S$ all of which are finite. 

Equation $(\ref{theta2})$ for $\cot\phi$ has an advantage over 
the original formulation \cite{Asada02b} for $\tan\phi$, 
\begin{equation}
\sum_{i=0}^5a_i(\tan\phi)^i=0 . 
\end{equation}
In particular, because $a_5$, the coefficient of $\tan^5\phi$, 
can vanish, $\tan\phi$ can be infinite, while, as seen from 
Eq. ($\ref{theta2}$), $\cot\phi$ remains finite, because 
$a_0$ never vanishes for off-axis sources. 
Hence, any solution of Eq. ($\ref{theta2}$) is finite. 
As a consequence, we can derive the following theorem: 
When the source moves around off the symmetry axis 
($\beta_y=0$, i.e., $S=0$), images never cross the separation 
axis ($\theta_y=0$, i.e., $\cot\phi=\pm\infty$). 
This implies that the number of images on each side 
of the separation axis is invariant, 
unless the source crosses the caustics. 
This property holds for arbitrary $\ell$ and $\nu$, 
because the number of images is an integer. 
An analytic study of a symmetric binary ($\nu=1/2$) 
\cite{SW,Asada02b} is sufficient to determine the invariant number. 
In this case, only a single image appears on the source side 
of the axis, while two images appear on the side opposite to 
the source outside the caustics or four images appear 
on the side opposite to the source inside the caustics. 

We also find the relation involving $r$ \cite{Asada02b} 
\begin{equation}
r \sin\phi=\frac{W_0(\cot\phi)}{W_1(\cot\phi)} , 
\label{rsin}
\end{equation}
where we defined 
\begin{eqnarray}
W_0(\cot\phi)&=&
(\ell^2\rho C-\ell\rho^2 C^2-\nu\ell+\rho C) 
\nonumber\\
&&-\rho S (\ell^2-2\ell\rho C+1)\cot\phi 
-\ell\rho^2 S^2 \cot^2\phi, \\
W_1(\cot\phi)&=&\rho (C-S\cot\phi)   
\Bigl[\rho S\cot^2\phi+2(\ell-\rho C)\cot\phi
-\rho S \Bigr] . 
\end{eqnarray}
This can be recast into the form 
\begin{equation}
r={\cal R}(\phi, \rho, \varphi, \ell, \nu) , 
\label{r}
\end{equation}
where ${\cal R}$ is a function of $\phi$, $\rho$, $\varphi$, 
$\ell$ and $\nu$. 

It should be noted that Eq. ($\ref{rsin}$) is obtained from 
the two equations 
\begin{eqnarray}
&&r^2-2\ell r\cos\phi+\ell^2=\frac{\nu\ell}
{\rho(C-S\cot\phi)} , 
\label{supplement}
\\
&&\Bigl[\ell-\rho(C-S\cot\phi)\Bigr]r^2\sin\phi 
-\ell\rho S r-\ell(1-\nu)\sin\phi=0 , 
\label{supplement2}
\end{eqnarray}
where $W_1 \neq 0$ has been assumed. 
Let us consider the case $W_1=0$ in detail. 
This means that $W_0=0$ also.  
Substituting Eq. ($\ref{supplement}$) into $r^2$ in 
Eq. ($\ref{supplement2}$) gives us the equation $W_0=0$ for 
$\cot\phi$, because the coefficients of $r$ vanish, due to $W_1=0$. 
For $W_0=W_1=0$, Eqs. ($\ref{supplement}$) and 
($\ref{supplement2}$) become identical. 
Now, we determine the condition for which there exists a $\cot\phi$ 
satisfying the two equations $W_0=0$ and $W_1=0$. 
It is given by the resultant \cite{Waerden} defined 
for $W_0=u_0\cot^2\phi+u_1\cot\phi+u_2$ and 
$W_1=v_0\cot^3\phi+v_1\cot^2\phi+v_2\cot\phi+v_3$ 
as the determinant  
\begin{equation}
W \equiv 
\left|
\begin{array}{ccccc}
u_0 & u_1 & u_2 & 0 & 0 \\
0 & u_0 & u_1 & u_2 & 0 \\
0 & 0 & u_0 & u_1 & u_2 \\
v_0 & v_1 & v_2 & v_3 & 0 \\
0 & v_0 & v_1 & v_2 & v_3 \\
\end{array}
\right| . 
\end{equation} 
Using Mathematica,\cite{Wolfram} this is evaluated as 
\begin{eqnarray}
W=S^2\rho^2&\Bigl[&
\ell^2\rho^4-4C\ell^3\rho^3-\rho^2(1-\ell^4-4C^2\ell^4-2\ell^2\nu)
\nonumber\\
&&+2C\ell\rho(1-\ell^4-2\ell^2\nu)-\ell^2\nu(2-2\ell^2-\nu) 
\Bigr] . 
\label{W}
\end{eqnarray}
If and only if $W=0$, there exists a $\cot\phi$ satisfying 
$W_0=W_1=0$. 
In this case, we must solve $W_0=W_1=0$ for $\cot\phi$. 
Then, substitution of the solution into Eq. ($\ref{supplement}$), 
which is identical to Eq. ($\ref{supplement2}$), gives us a quadratic 
equation in $r$. This can be solved immediately. 
This yields two solutions on a line defined by $y/x=\cot\phi$. 
We should note that $W=0$ is an artifact of our polar coordinates. 
In the following, we consider only the case $W\neq 0$. 

\subsection{Discriminant}
We factorized the discriminant \cite{Waerden} $D_5$ of 
the quintic equation Eq. ($\ref{theta2}$) using Mathematica 
\cite{Wolfram} into the form 
\begin{equation}
D_5=-4 S^2 \ell^2 \nu^2 \rho^2 W^2 K , 
\label{D5} 
\end{equation}
where we defined 
\begin{equation}
K=\sum^{16}_{i=0}k_i \rho^{16-i} , 
\end{equation}
for $k_0=\ell^8$, $k_1=8C\ell^7(1-\ell^2-2\nu)$,
$\cdots$. Here, $k_2, \cdots, k_{16}$ have rather lengthy forms.  

Changes in the number of images are usually studied in terms of 
the caustics ($J=0$). In our approach for a {\it real} algebraic 
equation, changes in the number of solutions for the lens equation 
Eq. ($\ref{theta2}$) correspond directly to those in the signature 
of the discriminant $D_5$. In the next subsection, we show 
that $K=0$ is a condition for the caustics.

\subsection{Caustics}
Caustics are curves on the source plane, where the Jacobian 
of the lens mapping 
$|\partial\mbox{\boldmath $\beta$}/\partial\mbox{\boldmath $\theta$}|$ 
vanishes. 
In our polar coordinates, the Jacobian is expressed as 
\begin{equation}
J \equiv 
\left| \frac{\partial \mbox{\boldmath $\beta$}}
{\partial\mbox{\boldmath $\theta$}} \right|
=\frac{\rho}{r}\frac{1}
{\frac{\partial (r, \phi)}{\partial (\rho, \varphi)}} . 
\end{equation}
Since $r$ and $\rho$ are nonvanishing and finite, 
$|\partial\mbox{\boldmath $\beta$}/\partial\mbox{\boldmath $\theta$}|$ 
vanishes where 
\begin{equation}
\frac{\partial (r, \phi)}{\partial (\rho, \varphi)}=\pm\infty . 
\end{equation}. 

Differentiation of Eq. ($\ref{theta2}$) gives 
\begin{equation}
\Phi d\phi-
\sum_{i=0}^5 (\cot\phi)^{5-i} 
\left(
\frac{\partial a_i}{\partial \rho}d\rho+
\frac{\partial a_i}{\partial \varphi}d\varphi
\right)=0 . 
\label{dphi}
\end{equation}
Here, we defined 
\begin{equation}
\Phi=\sum_{i=0}^6 b_i (\cot\phi)^{6-i} , 
\end{equation}
where $b_0=5a_0$, $b_1=4a_1$, $b_2=3a_2+5a_0$, 
$b_3=2a_3+4a_1$, $b_4=a_4+3a_2$, $b_5=2a_3$ and $b_6=a_4$.  
Therefore, we find 
\begin{eqnarray}
\frac{\partial \phi}{\partial \rho}
&=&\frac{1}{\Phi}
\sum_{i=0}^5 (\cot\phi)^{5-i}\frac{\partial a_i}{\partial \rho} ,
\\
\frac{\partial \phi}{\partial \varphi}
&=&\frac{1}{\Phi}
\sum_{i=0}^5 (\cot\phi)^{5-i}\frac{\partial a_i}{\partial \varphi} .
\end{eqnarray}

The total derivative of Equation ($\ref{r}$) leads to 
\begin{eqnarray}
\Phi dr&=&\left( \Phi\frac{\partial {\cal R}}{\partial \rho} 
+\frac{\partial {\cal R}}{\partial \phi} 
\sum_{i=0}^5 (\cot\phi)^{5-i}\frac{\partial a_i}{\partial \rho} 
\right) d\rho 
\nonumber\\
&&+\left( \Phi\frac{\partial {\cal R}}{\partial \varphi} 
+\frac{\partial {\cal R}}{\partial \phi} 
\sum_{i=0}^5 (\cot\phi)^{5-i}\frac{\partial a_i}{\partial \varphi} 
\right) d\varphi , 
\end{eqnarray}
where we used Eq. ($\ref{dphi}$) to eliminate $d\phi$. 
Hence, we obtain 
\begin{eqnarray}
\frac{\partial r}{\partial \rho}&=&
\frac{\partial {\cal R}}{\partial \rho}+
\frac{\partial {\cal R}}{\partial \phi}
\frac{\partial \phi}{\partial \rho} ,
\\
\frac{\partial r}{\partial \varphi}&=&
\frac{\partial {\cal R}}{\partial \varphi}+
\frac{\partial {\cal R}}{\partial \phi}
\frac{\partial \phi}{\partial \varphi} , 
\end{eqnarray}
so that we find the Jacobian in the polar coordinates as 
\begin{equation} 
\frac{\partial (r, \phi)}{\partial (\rho, \varphi)}
=\frac{\partial {\cal R}}{\partial \rho}
\frac{\partial \phi}{\partial \varphi}-
\frac{\partial {\cal R}}{\partial \varphi}  
\frac{\partial \phi}{\partial \rho} , 
\end{equation}
where $\partial {\cal R}/\partial \rho$ and 
$\partial {\cal R}/\partial \varphi$ are finite. 
The Jacobian $\partial (r, \phi)/\partial (\rho, \varphi)$ 
is infinite if and only if $\Phi=0$, because  
$\cot\phi$, $\partial a_i/\partial\rho$ and 
$\partial a_i/\partial\varphi$ in $\partial\phi/\partial\rho$ and 
$\partial\phi/\partial\varphi$ are all finite.  

The lens equation and the Jacobian set equal to $0$ are considered  
a set of equations for $\cot\phi$. 
The vanishing of the resultant \cite{Waerden} is the necessary and 
sufficient condition that there exists a $\cot\phi$ satisfying 
both equations. 
The resultant becomes 
\begin{equation}
J_5 \equiv 
\left|
\begin{array}{ccccccccccc}
a0 & a1 & a2 & a3 & a4 & a5 & 0 & 0 & 0 & 0 & 0 \\
0 & a0 & a1 & a2 & a3 & a4 & a5 & 0 & 0 & 0 & 0 \\
0 & 0 & a0 & a1 & a2 & a3 & a4 & a5 & 0 & 0 & 0 \\
0 & 0 & 0 & a0 & a1 & a2 & a3 & a4 & a5 & 0 & 0 \\
0 & 0 & 0 & 0 & a0 & a1 & a2 & a3 & a4 & a5 & 0 \\
0 & 0 & 0 & 0 & 0 & a0 & a1 & a2 & a3 & a4 & a5 \\ 
b0 & b1 & b2 & b3 & b4 & b5 & b6 & 0 & 0 & 0 & 0 \\
0 & b0 & b1 & b2 & b3 & b4 & b5 & b6 & 0 & 0 & 0 \\
0 & 0 & b0 & b1 & b2 & b3 & b4 & b5 & b6 & 0 & 0 \\
0 & 0 & 0 & b0 & b1 & b2 & b3 & b4 & b5 & b6 & 0 \\
0 & 0 & 0 & 0 & b0 & b1 & b2 & b3 & b4 & b5 & b6 \\
\end{array}
\right| , 
\end{equation}
which is a quite complicated function of 
$\nu$, $\ell$, $\rho$, $C$ and $S$. 
Using Mathematica,\cite{Wolfram} we factorized it into 
\begin{equation}
J_5=-64S^2\ell^4\nu^2(1-\nu)^2\rho^6 
\Bigl((\rho C-\ell)^2+\rho^2S^2\Bigr) W^2 K , 
\label{J5}
\end{equation}
where $(\rho C-\ell)^2+\rho^2S^2$ is always positive for off-axis
sources $(S \neq 0)$. 
Hence, from Eqs. ($\ref{D5}$) and ($\ref{J5}$), we have shown that 
$D_5=0$ corresponds to the caustics ($J_5=0$). 
Actually, $K=0$ is equivalent to the equation for the caustics 
obtained using the complex formalism.\cite{WP,Witt93}

\section{Conclusion}
Some algebraic properties of the real quintic equation for a binary 
gravitational lens have been studied. 
We find that the number of images on each side of the separation axis 
is independent of the mass ratio and separation 
unless the source crosses the caustics. 
This result should be useful for checking numerical results. 
Furthermore, the discriminant of the quintic equation was shown to 
factorize into two parts: One represents the condition that 
the lens equation can be reduced to a single quintic equation, 
while the other corresponds to the caustics. 

\section*{Acknowledgements}
We would like to thank M. Bartelmann for invaluable comments 
on the manuscript. 
H. A. would like to thank M. Bartelmann for his hospitality at the 
Max-Planck-Institut f\"ur Astrophysik, where a part of this work 
was done.
This work was supported in part by a Japanese Grant-in-Aid 
for Scientific Research from the Ministry of Education 
(No. 13740137) (H. A.) and the Sumitomo Foundation (H. A.). 


\end{document}